\documentclass[8.5pt,twoside,twocolumn,amsmath,amssymb]{article}
\usepackage[super,sort&compress,comma]{natbib}
\usepackage{mhchem}
\usepackage{times,mathptmx}
\usepackage{sectsty}
\usepackage{balance}

\usepackage{graphicx} 
\usepackage{lastpage}
\usepackage[format=plain,singlelinecheck=false,font=small,labelfont=bf,labelsep=space]{caption}
\usepackage{fancyhdr}
\usepackage{dcolumn}
\usepackage{bm}
\usepackage{amsmath}
\usepackage{bbding}
\usepackage{verbatim}
\usepackage{widetext}

\newcommand{\br}{{\bf r}}
\newcommand{\etal}{{\it et al.}}

\begin{document}

\thispagestyle{plain}
\fancypagestyle{plain}{
\fancyhead[L]{\includegraphics[height=8pt]{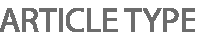}}
\fancyhead[C]{\hspace{-1cm}\includegraphics[height=20pt]{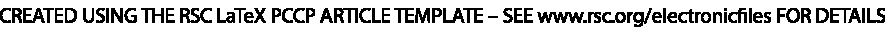}}
\fancyhead[R]{\includegraphics[height=10pt]{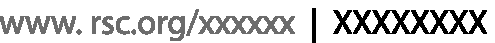}\vspace{-0.2cm}}
\renewcommand{\headrulewidth}{1pt}}
\renewcommand{\thefootnote}{\fnsymbol{footnote}}
\renewcommand\footnoterule{\vspace*{1pt}%
\hrule width 3.4in height 0.4pt \vspace*{5pt}}
\setcounter{secnumdepth}{5}

\makeatletter
\def\subsubsection{\@startsection{subsubsection}{3}{10pt}{-1.25ex plus -1ex minus -.1ex}{0ex plus 0ex}{\normalsize\bf}}
\def\paragraph{\@startsection{paragraph}{4}{10pt}{-1.25ex plus -1ex minus -.1ex}{0ex plus 0ex}{\normalsize\textit}}
\renewcommand\@biblabel[1]{#1}
\renewcommand\@makefntext[1]%
{\noindent\makebox[0pt][r]{\@thefnmark\,}#1}
\makeatother
\renewcommand{\figurename}{\small{Fig.}~}
\sectionfont{\large}
\subsectionfont{\normalsize}

\fancyfoot{}
\fancyfoot[LO,RE]{\vspace{-7pt}\includegraphics[height=9pt]{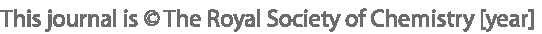}}
\fancyfoot[CO]{\vspace{-7.2pt}\hspace{12.2cm}\includegraphics{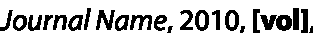}}
\fancyfoot[CE]{\vspace{-7.5pt}\hspace{-13.5cm}\includegraphics{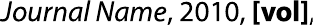}}
\fancyfoot[RO]{\footnotesize{\sffamily{1--\pageref{LastPage} ~\textbar  \hspace{2pt}\thepage}}}
\fancyfoot[LE]{\footnotesize{\sffamily{\thepage~\textbar\hspace{3.45cm} 1--\pageref{LastPage}}}}
\fancyhead{}
\renewcommand{\headrulewidth}{1pt}
\renewcommand{\footrulewidth}{1pt}
\setlength{\arrayrulewidth}{1pt}
\setlength{\columnsep}{6.5mm}
\setlength\bibsep{1pt}

\twocolumn[
  \begin{@twocolumnfalse}
\noindent\LARGE{\textbf{Ion Density Deviations in Semipermeable
Ionic Microcapsules}} \vspace{0.6cm}

\noindent\large{\textbf{Qiyun Tang\textit{$^{a,b}$} and Alan R. Denton\textit{$^{\ast}$\textit{$^{a}$}}
}}\vspace{0.5cm}

\noindent\textit{\small{\textbf{Received 15th February 2015, Accepted 21st March 2015}}}

\noindent \textbf{\small{DOI: 10.1039/c5cp00974j}}
\vspace{0.6cm}

\noindent \normalsize{By implementing the nonlinear Poisson-Boltzmann theory 
in a cell model, we theoretically investigate the influence of polyelectrolye 
gel permeability on ion densities and pH deviations inside the cavities 
of ionic microcapsules.  Our calculations show that variations in permeability 
of a charged capsule shell cause a redistribution of ion densities within 
the capsule, which ultimately affects the pH deviation and Donnan potential
induced by the electric field of the shell.  
We find that semipermeable capsules can induce larger pH deviations 
inside their cavities that can permeable capsules.  Furthermore, with increasing 
capsule charge, the influence of permeability on pH deviations progressively increases.  
Our theory, while providing a self-consistent method for modeling the influence 
of permeability on fundamental properties of ionic microgels, makes predictions 
of practical significance for the design of microcapsules loaded with 
fluorescent dyes, which can serve as biosensors for diagnostic purposes.} 
\vspace{0.5cm}
 \end{@twocolumnfalse}
  ]

\footnotetext{\textit{$^{a}$Department of Physics, North Dakota State University, Fargo, ND 58108-6050, USA.
E-mail: alan.denton@ndsu.edu}}
\footnotetext{\textit{$^{b}$Current address: Institut f{\"u}r Theoretische Physik,
Georg-August Universit{\"a}t, 37077 G{\"o}ttingen, Germany}}

\section{Introduction}

Polyelectrolyte (PE) microcapsules -- ionic (charged) colloidal particles 
with hollow cavities\cite{Vinogradova_2006_AnnuRevMaterRes,Borisov_2010_AFR}
-- have attracted great attention in the past decade due to their 
novel fundamental properties and their potential applications as 
biosensors to monitor local ion concentrations (such as pH) 
in cellular environments.\cite{Kuwana_2004_BP,Reibetanz_2010_BioM,delMercato_2011_Small,
Kreft_2007_JMC,DeGeest_2009_Softmatter,delMercato_2011_ACSNano,Xiaoxue_2014_JCIS,
Kazakova_2013_ABC,Sun_2012_BC,Dorleta_2012_CM,RiveraGil_2012_Small,
Lee_2011_Softmatter} Charged PE shells, which may encapsulate
pH-sensitive fluorescent dyes, generate electric fields that can cause
deviations in local ion distributions and limit practical applications. 
Previous experiments~\cite{Bostrom_2002_La,Zhang_2011_CPC} and
theories~\cite{Janata_1987_AC,Janata_1992_AC} demonstrated that the
pH and ion densities near a charged flat surface can deviate from their
bulk values.  For example, Bostrom \etal\cite{Bostrom_2002_La}
demonstrated that ion and pH gradients emerge near biological
flat membranes. Zhang \etal\cite{Zhang_2011_CPC} experimentally
confirmed that local ion concentrations and sensor read-outs can
be attributed to surface charges. Janata~\cite{Janata_1987_AC,Janata_1992_AC}
showed theoretically that pH shifts measured by optical sensors depend 
on bulk-surface interactions.
The charged shell of a spherical microcapsule can induce variations 
in ion densities near the strongly curved surface of the capsule. 
Furthermore, because of the asymmetry between the inside and outside 
environments, the measured ion concentrations in microcapsule cavities 
can deviate greatly from those in bulk. Understanding such ion deviations 
is significant for biomedical applications of ionic microcapsules 
as biosensors, e.g., to avoid misdiagnosis of diseases, such as early-stage
cancer.\cite{Weissleder_2006_Science,RiveraGil_2008_ACSNano,Xie_2011_ACR}
Ion distributions can also affect the release properties of pressurized
capsules for drug delivery.\cite{Schmidt_2011_AFM}

Ion density deviations inside of microcapsule cavities have been studied 
previously by experiments,\cite{Sukhorukov_1999_JPCB,Gao_2001_EPJE,
Halozan_2005_JCIM,Vinogradova_2005_Macromol} 
theories,\cite{Halozan_2005_JCIM,Halozan_2007_ACS} and simulations.\cite{Stukan_2006_PRE}
For example, Sukhorukov \etal\cite{Sukhorukov_1999_JPCB} and Gao \etal\cite{Gao_2001_EPJE}
demonstrated that a pH difference between the inside and outside of PE capsules 
emerges when the capsules are permeable to small ions, but exclude 
poly(styrenesulfonate) ions of a particular molecular weight.
Another approach, based on entrapping polyanions within PE microcapsules,
also reported a redistribution of H$^+$ ions across a semipermeable microcapsule
wall.\cite{Halozan_2005_JCIM} These ion density differences were attributed to 
the capsule semipermeability and explained by a macroscopic theoretical model 
of Donnan equilibrium.\cite{Halozan_2005_JCIM,Halozan_2007_ACS} 
Vinogradova \etal\cite{Vinogradova_2007_JCP,Vinogradova_2008_JCP,Lobaskin_2012_SM}
applied Poisson-Boltzmann (PB) theory to study the osmotic pressure 
acting on semipermeable shells in polyion solutions.
Stukan \etal\cite{Stukan_2006_PRE} used molecular dynamics simulation to study
ion distributions near nanocapsules that are permeable to solvent and counterions, 
but impermeable to polyelectrolyte coils, which were modeled as soft colloids.
Semipermeability to salt ions has also been linked to osmotic shock of 
rigid protein shells, such as viral capsids.\cite{Cordova_2003_BiophysJ} 

Recently, we demonstrated, using non-linear PB theory, 
that even fully permeable microcapsules can induce ion density deviations 
inside of charged capsule cavities.\cite{Tang_2014_PCCP} 
These deviations depend on the degree of dissociation of the PE making up 
the shells, rather than on the permeability of the capsule wall.
In practical applications of PE microcapsules in cellular environments, 
however, the capsule walls usually exclude polyions, such as charged DNA 
or amino acids. In this case, the ion density deviations induced by the 
microcapsule should also depend on the permeability of the charged wall.
The influence of permeability on deviations of local ion densities and pH 
induced by ionic microcapsules is still poorly understood.

In this paper, by modeling a PE microcapsule as a uniformly charged shell
that is permeable to some ionic species (counterions and salt ions), but 
only semipermeable to another species (polyions), we extend our previous work 
to analyze the impact of permeability of charged capsule shells on ion density
deviations in aqueous solutions. By employing nonlinear PB theory,
we systematically calculate deviations of local ion densities inside
microcapsule cavities induced by the permeability of capsule shells.
Unlike macroscopic models of Donnan equilibrium for neutral
capsules,\cite{Halozan_2005_JCIM,Halozan_2007_ACS} here the
redistribution of ion densities depends not only on the
capsule wall's permeability, but also on its degree of dissociation 
(charge density). Our results demonstrate that capsule permeability
can also significantly influence the properties of ionic microcapsules 
and their performance as biosensors.

The remainder of the paper is organized as follows.  In Sec.~\ref{models},
we define our model of semipermeable ionic microcapsules and describe our 
implementation of PB theory.  In Sec.~\ref{results}, we present numerical results 
from our calculations for the influence of capsule permeability on ion densities,
Donnan potentials, and pH distributions.  Finally, in Sec.~\ref{conclusions}, 
we summarize our findings and emphasize implications for practical applications.

\section{Models and Methods}\label{models}

\subsection{Uniform-Shell Model of Ionic Microcapsules}\label{capsule}

\begin{figure}
 \includegraphics[width=\columnwidth,angle=0]{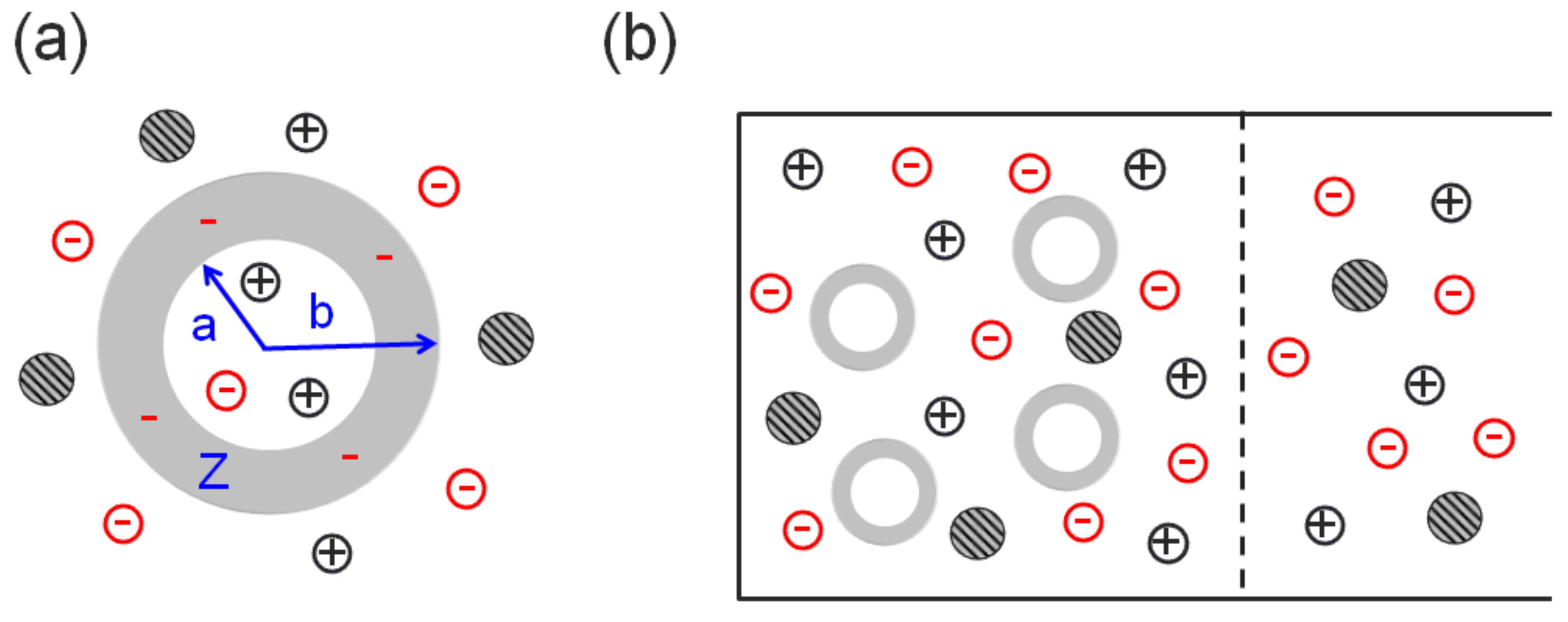}
\caption{ (a) Model of a spherical microcapsule of inner shell radius
$a$, outer shell radius $b$, and valence $Z$ in water.  The capsule
is permeable to water and small ions (smaller spheres labeled as $+$
and $-$), but only partially permeable to polyions (larger, shaded spheres).
(b) Bulk dispersion of microcapsules in Donnan equilibrium with a
reservoir.}\label{fig1}
\end{figure}

We consider a bulk dispersion of ionic capsules, microions, and polyions
dispersed in water. Within the primitive model of
polyelectrolytes,\cite{Oosawa_1971} the solvent is idealized as a
dielectric continuum with uniform dielectric constant $\epsilon$. We
model the capsules as spherical shells of inner radius $a$, outer
radius $b$, and valence $Z$.  In a polar solvent, a capsule becomes
charged when counterions dissociate from the PE chains that form the
cross-linked network (hydrogel) making up its shell.  Similarly, the
polyions become charged by dissociation. 
In line with many experiments, we presume the capsules 
to be negatively charged.  
Approximating the distribution of ionized sites in the hydrogel as uniform 
within the volume of the shell, we represent the number density of fixed charge
by a simple radial profile,
\begin{equation}
n_f(r)=\begin{cases} 0~,
&r>b~, \\[2ex]
{\displaystyle \frac{3Z}{4\pi(b^3-a^3)}}~,
&a<r<b~, \\[2ex]
0~, &0<r<a~, 
\end{cases}
\label{nf}
\end{equation}
where $r$ is radial distance from the center of the shell
[Fig.~\ref{fig1}(a)]. 
Although neglecting charge discreteness and ion-specific effects, 
this coarse-grained model is consistent with the primitive model 
and valid on length scales longer than the typical spacing 
between ionized groups on the PE backbones.

The microions comprise dissociated counterions and salt ions 
(e.g., Na$^+$ and Cl$^-$).  The polyion charge can be of either sign, 
depending on the species and the pH level of the solution.
For simplicity, we model microions and polyions as point charges.  
For reasons made clear below, we furthermore consider only 
monovalent ions.  

The capsules are permeable to water and microions, but only 
partially permeable to polyions. 
To quantify the degree of penetration of the capsule shell by polyions,
we introduce a permeability factor $\alpha$, which ranges from
0 (no penetration) to 1 (complete penetration). 
Within the cavity ($r<a$), we take the dielectric constant to be 
the same as in bulk, while in the shell, we assume a lower 
dielectric constant ($\epsilon_{\rm shell}<\epsilon$), 
as suggested by experiments on PE microgels in
water.\cite{Parthasarathy_1996_MSE,Mohanty_2012_SM} In Donnan
equilibrium, the capsules are confined to a fixed volume, while the
solvent, microions, and polyions can freely exchange with a
reservoir, with fixed number densities of salt ion pairs, $n_0$, and
of polyions, $n_{p0}$ [see Fig.~\ref{fig1}(b)].

\subsection{Poisson-Boltzmann Theory of Bulk Dispersions}\label{pb-theory}

We model a bulk dispersion of ionic capsules via Poisson-Boltzmann
theory.\cite{Deserno-Holm_2001} In its density-functional
formulation,\cite{Lowen_1992_PRL,Lowen_1993_JCP} PB theory focuses
on the grand potential functional, $\Omega[n_{\pm}(\br),
n_p(\br)]$, which is a unique functional of the number densities of
positive and negative microions, $n_{\pm}(\br)$, and
of polyions, $n_p(\br)$. 
We arbitrarily assume positively charged polyions of valence $z_p$.
In reduced form, the PB approximation for the grand potential 
functional can be expressed as
\begin{eqnarray}
&&\beta\Omega[n_{\pm}(\br), n_p(\br)]= \int d{\br}\, \left\{
n_+(\br)\left[\ln\left(\frac{n_+(\br)}{n_0}\right)-1\right] \right.
\nonumber\\[1ex]
&& \left.
+n_-(\br)\left[\ln\left(\frac{n_-(\br)}{n_0}\right)-1\right]
+n_p(\br)\left[\ln\left(\frac{n_p(\br)}{n_{p0}}\right)-1\right]
\right\}
\nonumber\\[1ex]
&& +\frac{1}{2}\int d{\br}\,
[n_+(\br)-n_-(\br)+z_pn_p(\br)-n_f(\br)]\psi(\br)~, \label{Omega}
\end{eqnarray}
where the first integral accounts for the ideal-gas free energy and
the second integral for the electrostatic potential energy.  Here,
$\beta\equiv 1/(k_BT)$ and $\psi(\br)\equiv\beta e\phi(\br)$ is a
dimensionless form of the electrostatic potential $\phi(\br)$,
\begin{equation}
e\phi(\br)=\int d{\br}'\, [n_+(\br)-n_-(\br)
+z_pn_p(\br)-n_f(\br)]v(|\br-\br'|)~, \label{phi}
\end{equation}
generated by Coulomb pair interactions, $v(r)=e^2/(\epsilon r)$,
between elementary charges $e$ on mobile microions dispersed
throughout the system and fixed ions localized within the shells.
The neglect of correlations among ions inherent in the
mean-field approximation for $\Omega$ is valid for weakly-correlated
monovalent ions, but questionable for more strongly correlated
multivalent microions.\cite{Vinogradova_2007_JCP,Vinogradova_2008_JCP,
Lobaskin_2012_SM}  For this reason, we restrict our considerations 
to monovalent ions.

In thermodynamic equilibrium, the grand potential functional 
is a minimum with respect to the microion and polyion densities.
Minimizing Eq.~(\ref{Omega}) with respect to $n_{\pm}({\bf r})$ and
$n_p({\bf r})$ yields Boltzmann distributions for the equilibrium
densities:
\begin{equation}
n_{\pm}({\bf r})=n_0 e^{\mp\psi({\bf r})} 
\label{nmu}
\end{equation}
and
\begin{equation}
n_p({\bf r})=
\begin{cases}
{\displaystyle n_{p0} e^{-z_p\psi({\bf r})}}~,
&r>b~, \\[2ex]
\alpha~n_{p0}e^{-z_p\psi({\bf r})}~, &0<r<b~,
\end{cases}
\label{ns}
\end{equation}
where the factor $\alpha$ dictates the permeability of the capsule 
to polyions, ranging from completely impermeable for $\alpha=0$
to completely permeable for $\alpha=1$.
Substituting Eqs.~(\ref{nmu}) and (\ref{ns}) into the Poisson equation,
\begin{equation}
\nabla^2\phi(\br)=-\frac{4\pi}{\epsilon}\rho(\br)~, \label{Poisson}
\end{equation}
where
\begin{equation}
\rho(\br)=e[n_+(\br)-n_-(\br)+z_pn_p(\br)-n_f(\br)] \label{rho}
\end{equation}
is the local charge density, yields the Poisson-Boltzmann equation:
\begin{equation}
\nabla^2\psi(\br)=\kappa_0^2\sinh\psi(\br)+4\pi\lambda_B
[n_f(\br)-z_pn_p(\br)]~. \label{PBeqn1}
\end{equation}
Here, $\lambda_B\equiv e^2/(\epsilon k_B T)$ is the Bjerrum length
($\lambda_B=0.714$ nm in water at temperature $T\simeq 293$,
$\epsilon=80$); $\kappa_0\equiv\sqrt{8\pi\lambda_Bn_0}$ is the
inverse Debye screening length in a reference reservoir of pure salt
solution (absent polyions); and
$\kappa_p\equiv\sqrt{4\pi\lambda_Bn_{p0}}$ can be interpreted as an
inverse Debye screening length of the polyions.  By solving
Eq.~(\ref{PBeqn1}) with appropriate boundary conditions, we obtain
the microion and polyion density profiles.

\vspace*{0.5cm}
\subsection{Poisson-Boltzmann Cell Model}\label{pb-cell-model}

In general, solving the PB equation [Eq.~(\ref{PBeqn1}] with boundary conditions 
matching an arbitrary arrangement of capsules is numerically quite challenging,
although feasible in some systems, such as charged colloids.\cite{Dobnikar_2003_JCP,
Dobnikar_2003_JPCM,Hallez_2014_Langmuir}  For computational efficiency, 
we instead use a cell model,\cite{Marcus_1955_JCP,Wennerstrom_1982_JCP} 
whose boundary conditions are relatively simple.  The cell model is justified 
by the wide disparity in size and charge between capsules and
microions and by our focus on solutions of relatively low salt
concentration.\cite{Hallez_2014_Langmuir,Denton_2010_JPCM}

For spherical capsules, the cell model represents a bulk dispersion 
by a spherical cell -- centered on a single capsule -- of radius $R$
determined by the capsule volume fraction $\eta=(b/R)^3$ 
[see Fig.~\ref{fig1}(a)].  Along with the capsule, the cell contains
counterions and salt ions, which may freely penetrate the capsule,
and polyions, whose penetration of the capsule is limited by 
the shell's permeability.
The condition of electroneutrality of the cell relates the ion numbers:
$Z=N_+-N_-+z_pN_p$.

In the spherical cell model, the PB equation simplifies to
\begin{widetext}
\begin{eqnarray}
\psi''(r)+\frac{2}{r}\psi'(r)= \begin{cases} 
{\displaystyle \kappa_0^2\sinh\psi(r)-z_p\kappa_p^2
e^{-z_p\psi(r)}}~, &b<r<R~, \\[2ex]
{\displaystyle \frac{1}{\chi}\left(\kappa_0^2\sinh\psi(r) -\alpha
z_p\kappa_p^2 e^{-z_p\psi(r)} +\frac{3Z\lambda_B}{b^3-a^3}\right)}~,
&a<r<b~, \\[2ex]
{\displaystyle \kappa_0^2 \sinh\psi(r)-\alpha z_p\kappa_p^2 e^{-z_p\psi(r)}}~,
&0<r<a~, 
\end{cases}\label{PBeqn2}
\end{eqnarray}
\end{widetext}
where $r$ is the radial distance from the center of the cell and
$\chi=\epsilon_{\rm shell}/\epsilon<1$ is the ratio of the
dielectric constant in the capsule shell to that in the bulk
solvent.

Boundary conditions on Eq.~(\ref{PBeqn2}) impose continuity of the
electrostatic potential at the inner and outer boundaries of the
capsule shell,
\begin{equation}
\psi_{\rm in}(a)=\psi_{\rm shell}(a)~, \hspace{1cm} \psi_{\rm
shell}(b)=\psi_{\rm out}(b)~, \label{boundP}
\end{equation}
vanishing of the electric field at the center of the cell and on the
cell boundary,
\begin{equation}
\psi'_{\rm in}(0)=0~, \hspace{1cm} \psi'_{\rm out}(R)=0~,
\label{boundE1}
\end{equation}
as required by spherical symmetry and electroneutrality, and
continuity of the electric displacement on the inner and outer shell
boundaries,
\begin{equation}
\psi'_{\rm in}(a)=\chi\psi'_{\rm shell}(a)~, \hspace{1cm}
\chi\psi'_{\rm shell}(b)=\psi'_{\rm out}(b)~, \label{boundE2}
\end{equation}
where the solutions in the three regions are labelled as $\psi_{\rm
in}(r)$ ($0<r<a$), $\psi_{\rm shell}(r)$ ($a<r<b$), and $\psi_{\rm
out}(r)$ ($b<r<R$).

By numerically solving the PB equation [Eq.~(\ref{PBeqn2})], along
with the boundary conditions [Eqs.~(\ref{boundP})-(\ref{boundE2})],
in the three radial regions (inside the cavity, in the shell, and
outside the capsule), we calculate the equilibrium microion and 
polyion density distributions within the spherical cell.

\section{Results and Discussion}\label{results}

\begin{figure}
 \includegraphics[width=\columnwidth,angle=0]{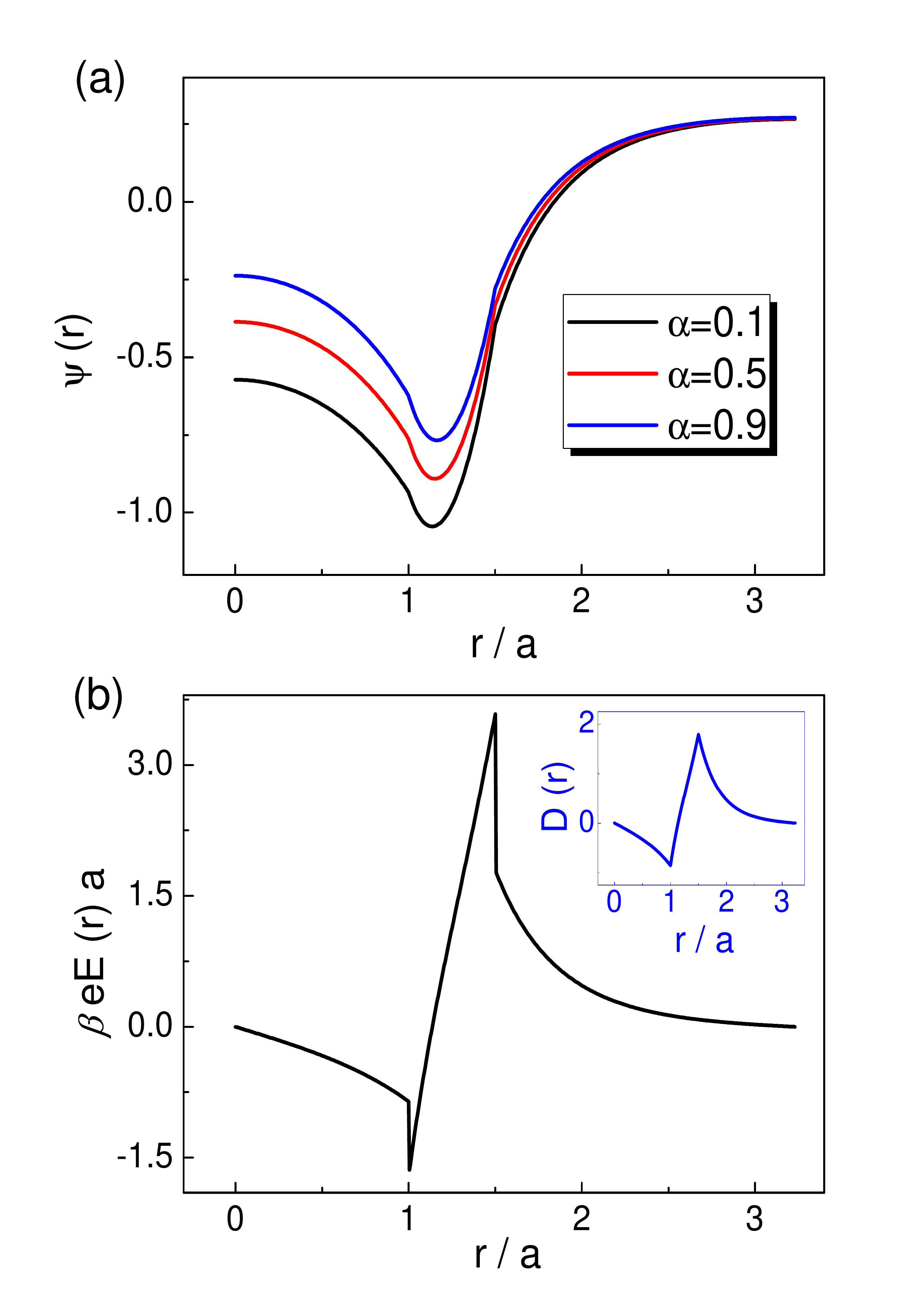}
  \caption{(a) Distribution of reduced electrostatic potential $\psi(r)$ 
  vs. radial distance $r$ from center of a microcapsule of inner shell radius 
  $a=50$ nm, outer radius $b=75$ nm, valence $Z=500$, and dielectric constant 
  ratio $\chi=0.5$ in an aqueous solution of microcapsule volume fraction 
  $\eta=0.1$, reservoir salt concentration $n_0=0.1$ mM, and polyion concentration 
  $n_p=0.04$ mM.
  Results are shown for shell permeability $\alpha=0.1$, 0.5, and 0.9, 
  (b) Electric field $E(r)$ vs. $r$ for permeability $\alpha=0.1$. 
  Inset shows distribution of displacement field $D(r)=\varepsilon E(r)$.
  }\label{fig2}
\end{figure}

\subsection{Distribution of Electrostatic Potential and Field}

To illustrate our theory, we present numerical results for system parameters 
typical of experiments on microcapsules.  Specifically, we consider 
negatively charged microcapsule shells 
of inner radius $a=50$ nm, outer radius $b=75$ nm, and valence $Z=500$ 
dispersed in water at room temperature ($\lambda_B=0.714$ nm), in Donnan equilibrium
with a salt reservoir of concentration $n_0=0.1$ mM, at volume fraction $\eta=0.1$,
corresponding to a cell radius $R=\eta^{-1/3}b\simeq 3.23~a$.
This concentration is sufficiently dilute to ensure independence of the 
ion distributions within neighboring cavities.  We set the dielectric constant
ratio between the microcapsule shell and the solution at $\chi=0.5$,
which is consistent with measured dielectric constants in hydrated ionic 
Poly(N-isopropylacrylamide) (PNIPAM) microgels ranging from 63 at 15$^{\circ}$C 
to 17 at 40$^{\circ}$C.\cite{Parthasarathy_1996_MSE,Mohanty_2012_SM}
In order to justify the mean-field PB approach, we consider here only 
weakly correlated monovalent polyions ($z_p=1$).
Finally, we choose the polyion concentration, $n_{p0}=0.4n_0$, to give
a polyion screening length $\kappa_p^{-1}$ somewhat longer than the
screening length $\kappa_0^{-1}$ in a reference reservoir of pure salt solution.

Figure~\ref{fig2}(a) shows the distribution of electrostatic potential 
within the cell.  Starting from the center, the potential decreases 
with increasing radial distance, reaching a minimum within the shell.
The depth of this minimum increases with decreasing capsule permeability,
reaching a value of $\sim$1 $k_BT$ (in potential energy) for $\alpha=0.1$.
The kinetic barrier to thermal diffusion of ions across the shell is thus 
sufficiently low to ensure equilibrium ion distributions.  Outside the shell,
the electrostatic potential increases toward the cell edge.  

The corresponding electric field is shown in Fig.~\ref{fig2}(b).  
Starting from zero at the center, the field decreases 
as $r$ increases inside the cavity, but then rises sharply within the shell.  
The discontinuities at the shell boundaries ($r=a$ and 1.5$a$) originate from 
the difference in dielectric constant between the shell and the solution. 
The displacement field, rather than the electric field, is continuous
at the boundaries [see inset to Fig.~\ref{fig2}(b)].  Outside the shell, 
the field decreases as $r$ increases, approaching zero at the edge of the cell 
to match the outer boundary condition. 
As a consistency check on our implementation of the theoretical model, 
we calculate the total charge number $N_{tot}$ of microions and polyions
in the cell,
\begin{equation}
N_{tot}=4\pi\int_0^R dr\, r^{2} [n_+(r)-n_-(r)+z_pn_p(r)]~,
\end{equation}
and confirm global electroneutrality ($N_{tot}=Z$) over a range of permeabilities.

\subsection{Donnan Potentials}

\begin{figure}
 \includegraphics[width=\columnwidth,angle=0]{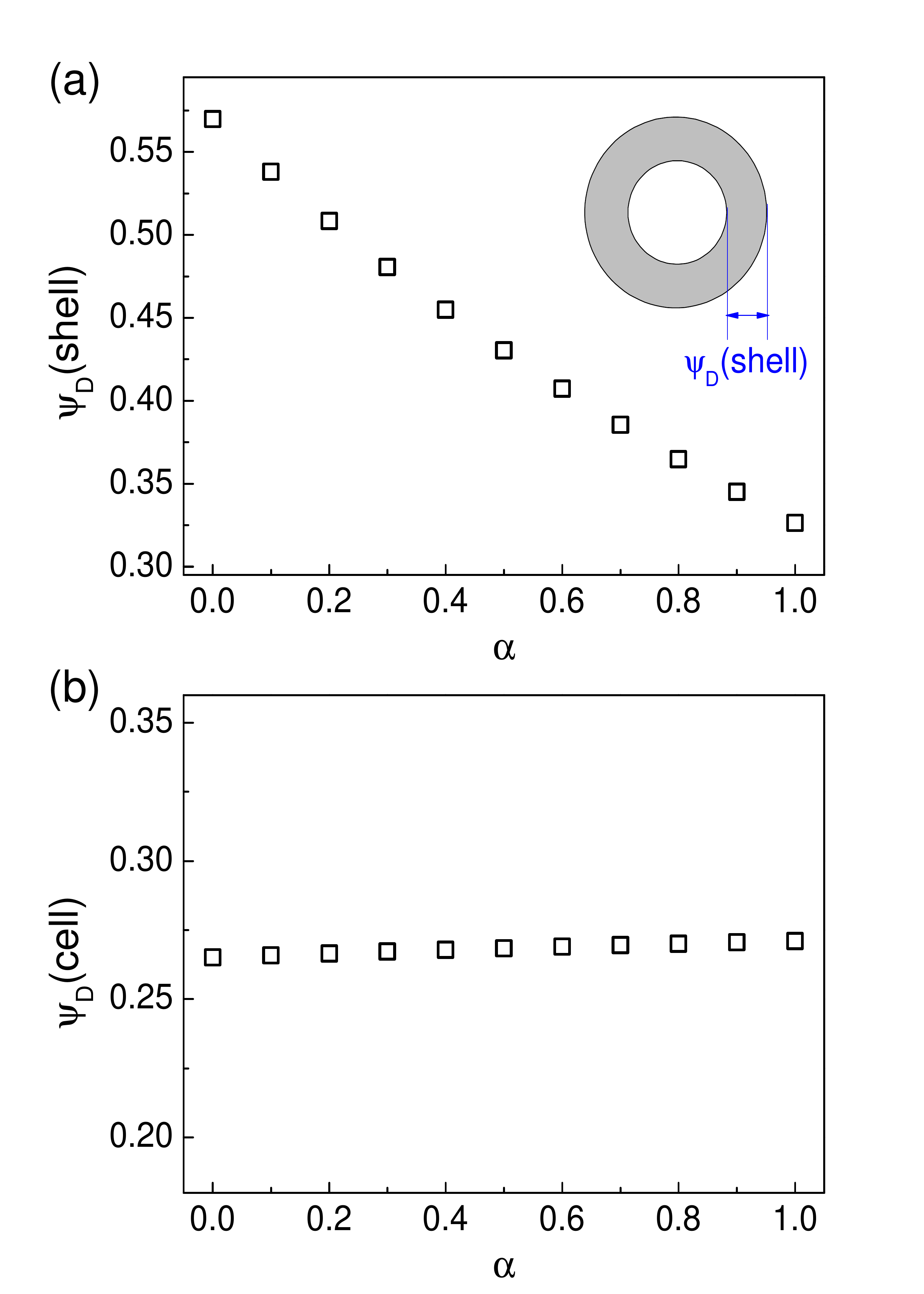}
  \caption{Donnan potential $\psi_D$ vs. capsule permeability $\alpha$
  at the edge of (a) the capsule shell ($r=a$) and (b) the entire cell ($r=R$).
  Other system parameters are as in Fig.~\ref{fig2}.
  }\label{fig3}
\end{figure}

Chemical equilibrium between the inside and outside environments of ionic
microcapsules can result in a significant difference in electrostatic potential 
between the two sides of the shell, as revealed in Fig.~\ref{fig2}(a).
We define this difference as the {\it shell Donnan potential}:
\begin{equation}
\psi_{D}({\rm shell})\equiv\psi(b)-\psi(a)~,
\end{equation}
which is analogous to the Donnan potential at the surface of a bulk
polyelectrolyte gel.\cite{Flory1953}
The variation of the shell Donnan potential with capsule permeability
$\alpha$ is shown in Fig.~\ref{fig3}(a).  With increasing permeability,
polyions increasingly penetrate the capsule shell, leading to
a gradual decrease of $\psi_{D}({\rm shell})$.  For comparison, 
Fig.~\ref{fig3}(b) shows the Donnan potential at the cell edge,
$\psi_{D}({\rm cell})\equiv\psi(R)$, as a function of capsule permeability. 
Interestingly, $\psi_{D}({\rm cell})$ is relatively insensitive to 
variation of $\alpha$, implying that capsule permeability also does not 
significantly influence the bulk osmotic pressure of the dispersion.
On the other hand, the permeability {\it is} expected, to influence 
ion distributions inside and outside the capsules, with potential impact 
on practical applications of ionic microcapsules, such as in diagnosis
of diseases.
In the remainder of this section, we examine in detail ion density 
distributions induced by shell permeability,
as well as corresponding pH deviations inside microcapsule cavities.

\subsection{Influence of Permeability on Ion Densities}\label{densities}
\begin{figure}
 \includegraphics[width=\columnwidth,angle=0]{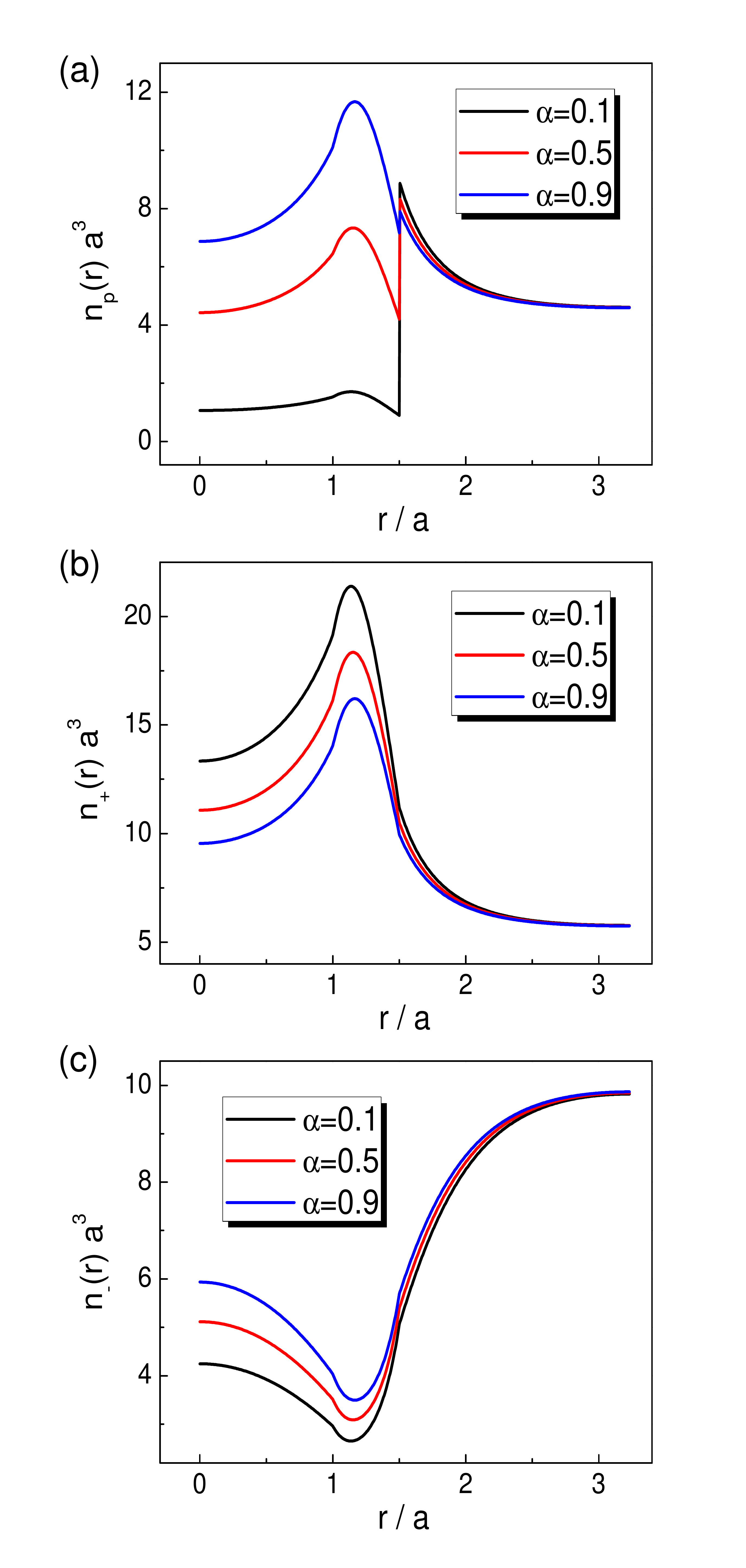}
  \caption{Density distributions of (a) polyions, (b) positive microions,
  and (c) negative microions vs. radial distance $r$ from center of 
  microcapsule for permeability $\alpha=0.1$, $0.5$, and $0.9$.
  Other system parameters are as in Fig.~\ref{fig2}.
  }\label{fig4}
\end{figure}

Next, we investigate the influence of capsule permeability on the ion
density distributions within the cell.  In practice, permeability may increase 
upon swelling, triggered by a change of solution pH or temperature.
Figure~\ref{fig4}(a) shows the distribution of polyion density 
over a range of permeabilities.  As the capsule becomes more permeable
($\alpha$ increases), polyions can more easily penetrate the capsule shell,
thus increasing the polyion density within the cavity.  This redistribution 
of polyions results from entropy-driven thermal diffusion, which can overcome 
the electrostatic potential energy barrier [see Fig.~\ref{fig2}(a)].

The rising concentration of polyions inside the capsule upon increasing 
permeability leads to an expulsion of positive microions and a corresponding
decrease in $n_+(r)$ for $r<b$, as shown in Fig.~\ref{fig4}(b).  Nevertheless, 
there is a net increase in the total concentration of positive ions inside
the capsule.  To maintain electroneutrality of the system, the density of 
negative microions also increases, as shown in Fig.~\ref{fig4}(c). 
The redistributions of ion densities induced by changes in capsule
permeability should be mirrored by corresponding changes in the local 
concentrations of H$^+$ and OH$^-$ ions, leading to deviations in local pH
from bulk solution values.  

\subsection{Influence of Permeability on pH in Microcapsules}

\begin{figure}
 \includegraphics[width=\columnwidth,angle=0]{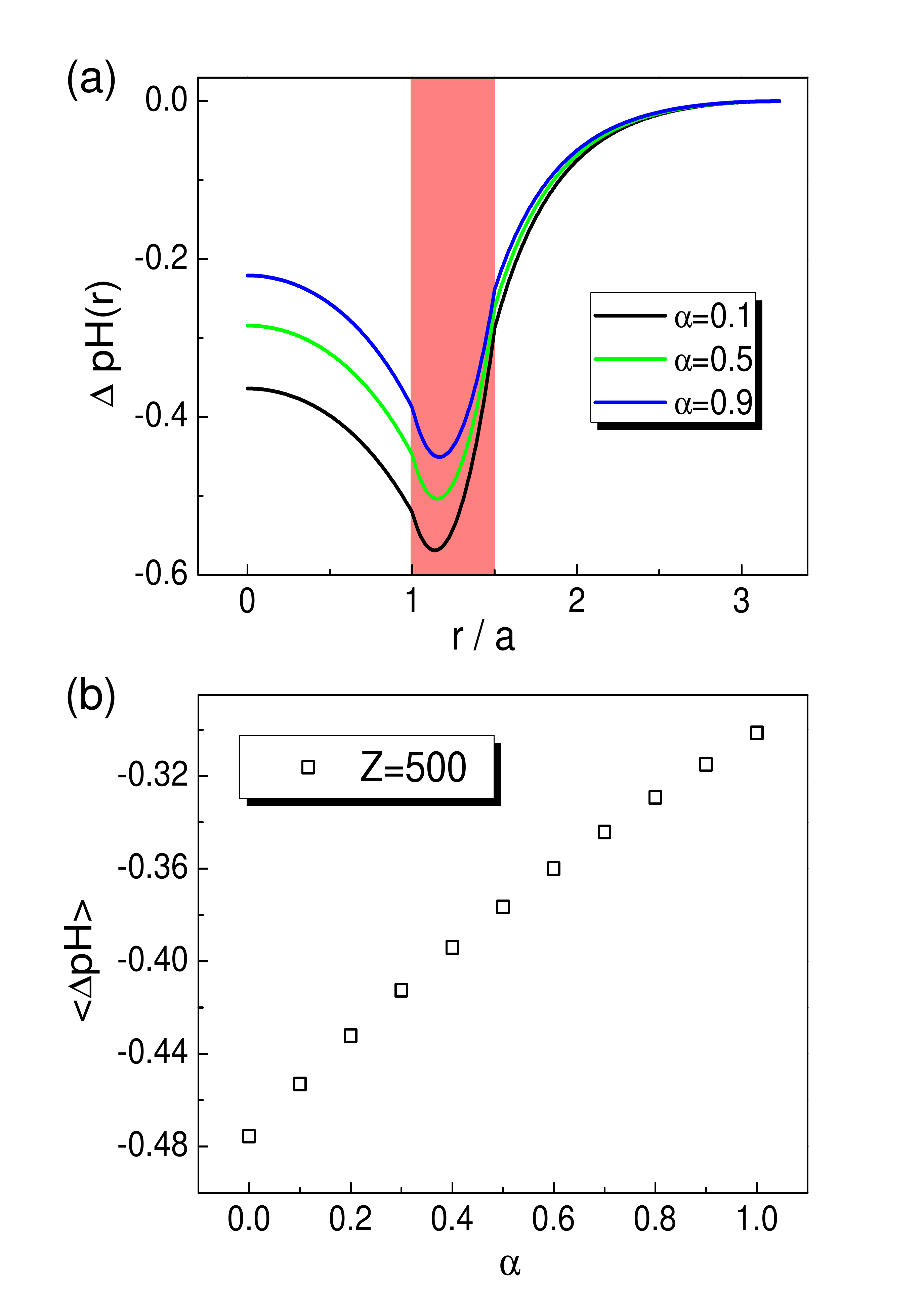}
  \caption{(a) Deviation of local pH from bulk value vs. radial distance $r$
  from center of microcapsule [Eq.~(\ref{delta-pH})] for valence $Z=500$ 
  and permeability $\alpha=0.1$, $0.5$, and $0.9$.
  (b) Deviation of average pH inside cavity ($r<a$) vs. permeability [Eq.~(\ref{pHav})].
  Other system parameters are as in Fig.~\ref{fig2}.
  }\label{fig5}
\end{figure}

Next, we investigate the influence of microcapsule permeability on pH 
deviations inside the capsule cavities and explore the potential impact 
on applications to pH sensors in cellular environments.
The polyions in our model are positively charged, corresponding to 
the real scenario of amino acids dispersed in an alkaline cellular environment.  
Under such conditions, the pH within the cell is determined by the concentration 
of hydroxyl (OH$^-$) ions.  Assuming the concentration of OH$^-$ ions to be 
proportional to that of all negative microions, the local pH deviation 
induced by the charged shell can be approximated by
\begin{equation}
\Delta{\rm pH}(r)=\log[n_{-}(r)/n_-(R)] \label{delta-pH}~.
\end{equation}
The average pH deviation inside the capsule cavities is then
\begin{equation}
\langle\Delta{\rm pH}\rangle\equiv\frac{3}{a^3}\int_0^a dr\, r^2
\Delta{\rm pH}(r)~. \label{pHav}
\end{equation}
For a capsule of valence $Z=500$, Fig.~\ref{fig5} shows (a) the local pH 
deviation profile and (b) the average pH deviation induced by the charged 
shell over a range of capsule permeabilities.  With increasing $\alpha$, 
the local and average pH deviations inside the cavity decrease in magnitude,
as a result of variations in ion concentrations discussed
in Sec.~\ref{densities}.  Thus, in alkaline environments, pH deviations 
inside cavities of ionic microcapsules induced by the charged shells are 
suppressed by increasing permeability to positive polyions.
In other words, semipermeable charged capsules can induce larger pH deviations 
inside their cavities that can permeable capsules.

\begin{figure}
 \includegraphics[width=\columnwidth,angle=0]{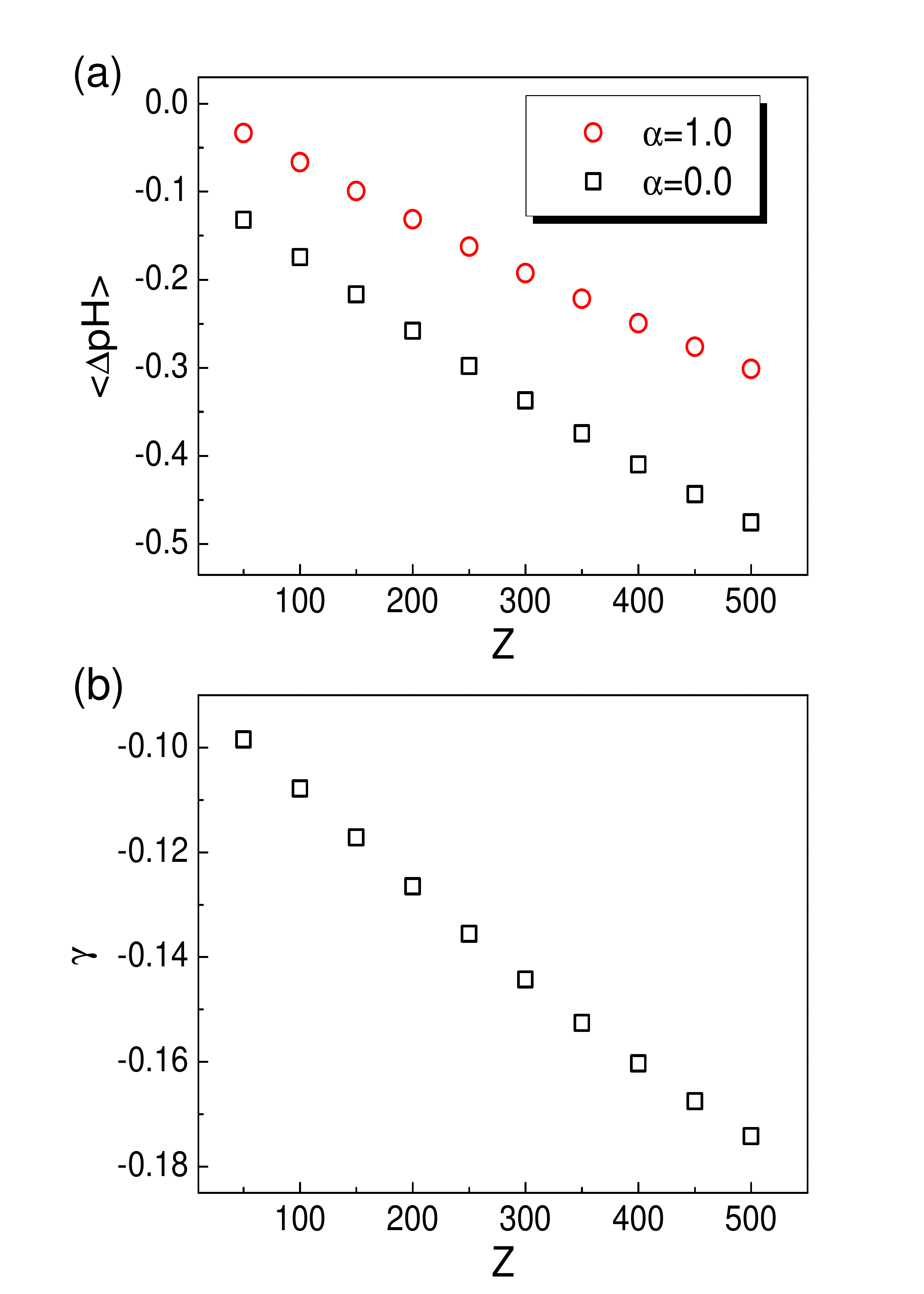}
  \caption{(a) Average deviation of pH from bulk value inside microcapsule 
  cavity vs. valence $Z$ for permeability $\alpha=0$ and $1$.
  (b) Permeability-induced average pH deviation vs. valence [Eq.~(\ref{gamma})].
  Other system parameters are as in Fig.~\ref{fig2}.
  }\label{fig6}
\end{figure}

In a previous paper,\cite{Tang_2014_PCCP} we demonstrated that the 
valence of ionic microcapsules dominates pH deviations inside the cavities.
The larger the capsule valence, the greater the pH deviations. 
Here we find that the permeability of the capsule can also influence 
pH deviations.  Finally, we investigate how the deviations induced by 
permeability vary with capsule valence.  Figure~\ref{fig6}(a) illustrates 
the variation with capsule valence of the average pH deviation inside the 
cavity for the two extremes of permeability, $\alpha=0$ and $1$. 
We find that, with increasing valence, the average pH deviation increases
in magnitude.  For capsules that are impermeable to polyions ($\alpha=0$),
however, the increase is more rapid than for fully permeable capsules ($\alpha=1$).
To quantify this effect, we define 
\begin{equation}
\gamma=\langle\Delta{\rm pH}\rangle_{\alpha=0}-\langle\Delta{\rm pH}
\rangle_{\alpha=1}\label{gamma}~,
\end{equation}
which measures the average pH deviation inside the capsule cavity induced 
solely by permeability of the capsule shell.
Figure~\ref{fig6}(b) shows the permeability-induced average pH deviation 
inside the cavities as a function of capsule valence, illustrating that,
as valence increases, the average pH deviation associated with capsule permeability 
increases in magnitude.

\section{Conclusions}\label{conclusions}
In summary, by applying a cell model implementation of the nonlinear 
Poisson-Boltzmann theory to solutions of ionic microcapsules that are
semipermeable to polyions, we have analyzed the influence of capsule 
permeability on ion densities and pH deviations inside the capsule cavities.
In real biological systems, variations in microcapsule permeability could 
result from variations in porosity of the polyelectrolyte networks making up 
the capsule shells associated with the size distribution of polyions 
(e.g., amino acids) or from network swelling/de-swelling induced by 
changes in temperature, pH, or salt concentration. 

Our results show that, upon varying capsule permeability, the ion densities 
redistribute so as to fulfill the competing requirements of minimum free energy 
and global electroneutrality.  Ultimately, increasing permeability suppresses 
deviations in microion density and pH induced by the charged shells.
These findings have potential relevance for the design of microcapsules that 
encapsulate fluorescent dyes to serve as ionic biosensors for diagnostic purposes.
Although we have focused, in this study, on solutions containing only
monovalent ions, in order to justify our use of the mean-field PB theory,
our approach could be extended to multivalent ion solutions by incorporating
ion correlations into the PB theory~\cite{Forsman_2004_JPCB}
or by performing molecular simulations in the cell model.

\vspace*{1cm}
\noindent{\bf \large Acknowledgments} \\[1ex]
This work was supported by the National Science Foundation under
Grant No. DMR-1106331.

\bibliographystyle{rsc}

\begin{mcitethebibliography}{47}
\providecommand*{\natexlab}[1]{#1}
\providecommand*{\mciteSetBstSublistMode}[1]{}
\providecommand*{\mciteSetBstMaxWidthForm}[2]{}
\providecommand*{\mciteBstWouldAddEndPuncttrue}
  {\def\EndOfBibitem{\unskip.}}
\providecommand*{\mciteBstWouldAddEndPunctfalse}
  {\let\EndOfBibitem\relax}
\providecommand*{\mciteSetBstMidEndSepPunct}[3]{}
\providecommand*{\mciteSetBstSublistLabelBeginEnd}[3]{}
\providecommand*{\EndOfBibitem}{}
\mciteSetBstSublistMode{f}
\mciteSetBstMaxWidthForm{subitem}
{(\emph{\alph{mcitesubitemcount}})}
\mciteSetBstSublistLabelBeginEnd{\mcitemaxwidthsubitemform\space}
{\relax}{\relax}

\bibitem[Vinogradova \emph{et~al.}(2006)Vinogradova, Lebedeva, and
  Kim]{Vinogradova_2006_AnnuRevMaterRes}
O.~I. Vinogradova, O.~V. Lebedeva and B.-S. Kim, \emph{Annu. Rev. Mater. Res.},
  2006, \textbf{36}, 143--78\relax
\mciteBstWouldAddEndPuncttrue
\mciteSetBstMidEndSepPunct{\mcitedefaultmidpunct}
{\mcitedefaultendpunct}{\mcitedefaultseppunct}\relax
\EndOfBibitem
\bibitem[Borisov \emph{et~al.}(2010)Borisov, Mayr, Mistlberger, and
  Klimant]{Borisov_2010_AFR}
S.~M. Borisov, T.~Mayr, G.~Mistlberger and I.~Klimant, in \emph{Advanced
  Fluorescence Reporters in Chemistry and Biology II: Molecular Constructions,
  Polymers and Nanoparticles}, ed. A.~P. Demchenko, 2010, vol.~09, pp.
  193--228\relax
\mciteBstWouldAddEndPuncttrue
\mciteSetBstMidEndSepPunct{\mcitedefaultmidpunct}
{\mcitedefaultendpunct}{\mcitedefaultseppunct}\relax
\EndOfBibitem
\bibitem[Kuwana \emph{et~al.}(2004)Kuwana, Liang, and
  Sevick-Muraca]{Kuwana_2004_BP}
E.~Kuwana, F.~Liang and E.~M. Sevick-Muraca, \emph{Biotechnol. Prog.}, 2004,
  \textbf{20}, 1561--1566\relax
\mciteBstWouldAddEndPuncttrue
\mciteSetBstMidEndSepPunct{\mcitedefaultmidpunct}
{\mcitedefaultendpunct}{\mcitedefaultseppunct}\relax
\EndOfBibitem
\bibitem[Reibetanz \emph{et~al.}(2010)Reibetanz, Chen, Mutukumaraswamy, Liaw,
  Oh, Venkatraman, Donath, and Neu]{Reibetanz_2010_BioM}
U.~Reibetanz, M.~H.~A. Chen, S.~Mutukumaraswamy, Z.~Y. Liaw, B.~H.~L. Oh,
  S.~Venkatraman, E.~Donath and B.~Neu, \emph{Biomacromolecules}, 2010,
  \textbf{11}, 1779--1784\relax
\mciteBstWouldAddEndPuncttrue
\mciteSetBstMidEndSepPunct{\mcitedefaultmidpunct}
{\mcitedefaultendpunct}{\mcitedefaultseppunct}\relax
\EndOfBibitem
\bibitem[del Mercato \emph{et~al.}(2011)del Mercato, Abbasi, and
  Parak]{delMercato_2011_Small}
L.~L. del Mercato, A.~Z. Abbasi and W.~J. Parak, \emph{Small}, 2011,
  \textbf{7}, 351--363\relax
\mciteBstWouldAddEndPuncttrue
\mciteSetBstMidEndSepPunct{\mcitedefaultmidpunct}
{\mcitedefaultendpunct}{\mcitedefaultseppunct}\relax
\EndOfBibitem
\bibitem[Kreft \emph{et~al.}(2007)Kreft, Javier, Sukhorukov, and
  Parak]{Kreft_2007_JMC}
O.~Kreft, A.~M. Javier, G.~B. Sukhorukov and W.~J. Parak, \emph{J. Mater.
  Chem.}, 2007, \textbf{17}, 4471--4476\relax
\mciteBstWouldAddEndPuncttrue
\mciteSetBstMidEndSepPunct{\mcitedefaultmidpunct}
{\mcitedefaultendpunct}{\mcitedefaultseppunct}\relax
\EndOfBibitem
\bibitem[De~Geest \emph{et~al.}(2009)De~Geest, De~Koker, Sukhorukov, Kreft,
  Parak, Skirtach, Demeester, De~Smedt, and Hennink]{DeGeest_2009_Softmatter}
B.~G. De~Geest, S.~De~Koker, G.~B. Sukhorukov, O.~Kreft, W.~J. Parak, A.~G.
  Skirtach, J.~Demeester, S.~C. De~Smedt and W.~E. Hennink, \emph{Soft Matter},
  2009, \textbf{5}, 282--291\relax
\mciteBstWouldAddEndPuncttrue
\mciteSetBstMidEndSepPunct{\mcitedefaultmidpunct}
{\mcitedefaultendpunct}{\mcitedefaultseppunct}\relax
\EndOfBibitem
\bibitem[del Mercato \emph{et~al.}(2011)del Mercato, Abbasi, Ochs, and
  Parak]{delMercato_2011_ACSNano}
L.~L. del Mercato, A.~Z. Abbasi, M.~Ochs and W.~J. Parak, \emph{ACS Nano},
  2011, \textbf{5}, 9668--9674\relax
\mciteBstWouldAddEndPuncttrue
\mciteSetBstMidEndSepPunct{\mcitedefaultmidpunct}
{\mcitedefaultendpunct}{\mcitedefaultseppunct}\relax
\EndOfBibitem
\bibitem[Song \emph{et~al.}(2014)Song, Li, Tong, and Gao]{Xiaoxue_2014_JCIS}
X.~Song, H.~Li, W.~Tong and C.~Gao, \emph{J. Colloid Interface Sci.}, 2014,
  \textbf{416}, 252--257\relax
\mciteBstWouldAddEndPuncttrue
\mciteSetBstMidEndSepPunct{\mcitedefaultmidpunct}
{\mcitedefaultendpunct}{\mcitedefaultseppunct}\relax
\EndOfBibitem
\bibitem[Kazakova \emph{et~al.}(2013)Kazakova, Shabarchina, Anastasova, Pavlov,
  Vadgama, Skirtach, and Sukhorukov]{Kazakova_2013_ABC}
L.~I. Kazakova, L.~I. Shabarchina, S.~Anastasova, A.~M. Pavlov, P.~Vadgama,
  A.~G. Skirtach and G.~B. Sukhorukov, \emph{Anal. Bioanal. Chem.}, 2013,
  \textbf{405}, 1559--1568\relax
\mciteBstWouldAddEndPuncttrue
\mciteSetBstMidEndSepPunct{\mcitedefaultmidpunct}
{\mcitedefaultendpunct}{\mcitedefaultseppunct}\relax
\EndOfBibitem
\bibitem[Sun \emph{et~al.}(2012)Sun, Benjaminsen, Almdal, and
  Andresen]{Sun_2012_BC}
H.~Sun, R.~V. Benjaminsen, K.~Almdal and T.~L. Andresen, \emph{Bioconjugate
  Chem.}, 2012, \textbf{23}, 2247--2255\relax
\mciteBstWouldAddEndPuncttrue
\mciteSetBstMidEndSepPunct{\mcitedefaultmidpunct}
{\mcitedefaultendpunct}{\mcitedefaultseppunct}\relax
\EndOfBibitem
\bibitem[Jimenez~de Aberasturi \emph{et~al.}(2012)Jimenez~de Aberasturi,
  Montenegro, Ruiz~de Larramendi, Rojo, Klar, Alvarez-Puebla, Liz-Marzan, and
  Parak]{Dorleta_2012_CM}
D.~Jimenez~de Aberasturi, J.-M. Montenegro, I.~Ruiz~de Larramendi, T.~Rojo,
  T.~A. Klar, R.~Alvarez-Puebla, L.~M. Liz-Marzan and W.~J. Parak, \emph{Chem.
  Mater.}, 2012, \textbf{24}, 738--745\relax
\mciteBstWouldAddEndPuncttrue
\mciteSetBstMidEndSepPunct{\mcitedefaultmidpunct}
{\mcitedefaultendpunct}{\mcitedefaultseppunct}\relax
\EndOfBibitem
\bibitem[Rivera-Gil \emph{et~al.}(2012)Rivera-Gil, Nazarenus, Ashraf, and
  Parak]{RiveraGil_2012_Small}
P.~Rivera-Gil, M.~Nazarenus, S.~Ashraf and W.~J. Parak, \emph{Small}, 2012,
  \textbf{8}, 943--948\relax
\mciteBstWouldAddEndPuncttrue
\mciteSetBstMidEndSepPunct{\mcitedefaultmidpunct}
{\mcitedefaultendpunct}{\mcitedefaultseppunct}\relax
\EndOfBibitem
\bibitem[Lee \emph{et~al.}(2011)Lee, Tiwari, and Raghavan]{Lee_2011_Softmatter}
H.-Y. Lee, K.~R. Tiwari and S.~R. Raghavan, \emph{Soft Matter}, 2011,
  \textbf{7}, 3273--3276\relax
\mciteBstWouldAddEndPuncttrue
\mciteSetBstMidEndSepPunct{\mcitedefaultmidpunct}
{\mcitedefaultendpunct}{\mcitedefaultseppunct}\relax
\EndOfBibitem
\bibitem[Bostrom \emph{et~al.}(2002)Bostrom, Williams, and
  Ninham]{Bostrom_2002_La}
M.~Bostrom, D.~R.~M. Williams and B.~W. Ninham, \emph{Langmuir}, 2002,
  \textbf{18}, 8609--8615\relax
\mciteBstWouldAddEndPuncttrue
\mciteSetBstMidEndSepPunct{\mcitedefaultmidpunct}
{\mcitedefaultendpunct}{\mcitedefaultseppunct}\relax
\EndOfBibitem
\bibitem[Zhang \emph{et~al.}(2010)Zhang, Ali, Amin, Feltz, Oheim, and
  Parak]{Zhang_2011_CPC}
F.~Zhang, Z.~Ali, F.~Amin, A.~Feltz, M.~Oheim and W.~J. Parak, \emph{Chem.
  Phys. Chem.}, 2010, \textbf{11}, 730--735\relax
\mciteBstWouldAddEndPuncttrue
\mciteSetBstMidEndSepPunct{\mcitedefaultmidpunct}
{\mcitedefaultendpunct}{\mcitedefaultseppunct}\relax
\EndOfBibitem
\bibitem[Janata(1987)]{Janata_1987_AC}
J.~Janata, \emph{Anal. Chem.}, 1987, \textbf{59}, 1351--1356\relax
\mciteBstWouldAddEndPuncttrue
\mciteSetBstMidEndSepPunct{\mcitedefaultmidpunct}
{\mcitedefaultendpunct}{\mcitedefaultseppunct}\relax
\EndOfBibitem
\bibitem[Janata(1992)]{Janata_1992_AC}
J.~Janata, \emph{Anal. Chem.}, 1992, \textbf{64}, 921A--927A\relax
\mciteBstWouldAddEndPuncttrue
\mciteSetBstMidEndSepPunct{\mcitedefaultmidpunct}
{\mcitedefaultendpunct}{\mcitedefaultseppunct}\relax
\EndOfBibitem
\bibitem[Weissleder(2006)]{Weissleder_2006_Science}
R.~Weissleder, \emph{Science}, 2006, \textbf{312}, 1168--1171\relax
\mciteBstWouldAddEndPuncttrue
\mciteSetBstMidEndSepPunct{\mcitedefaultmidpunct}
{\mcitedefaultendpunct}{\mcitedefaultseppunct}\relax
\EndOfBibitem
\bibitem[Rivera~Gil and Parak(2008)]{RiveraGil_2008_ACSNano}
P.~Rivera~Gil and W.~J. Parak, \emph{ACS Nano}, 2008, \textbf{2},
  2200--2205\relax
\mciteBstWouldAddEndPuncttrue
\mciteSetBstMidEndSepPunct{\mcitedefaultmidpunct}
{\mcitedefaultendpunct}{\mcitedefaultseppunct}\relax
\EndOfBibitem
\bibitem[Xie \emph{et~al.}(2011)Xie, Liu, Eden, Ai, and Chen]{Xie_2011_ACR}
J.~Xie, G.~Liu, H.~S. Eden, H.~Ai and X.~Chen, \emph{Acc. Chem. Res.}, 2011,
  \textbf{44}, 883--892\relax
\mciteBstWouldAddEndPuncttrue
\mciteSetBstMidEndSepPunct{\mcitedefaultmidpunct}
{\mcitedefaultendpunct}{\mcitedefaultseppunct}\relax
\EndOfBibitem
\bibitem[Schmidt \emph{et~al.}(2011)Schmidt, Fernandes, De~Geest, Delcea,
  M{\"o}hwald, and Fery]{Schmidt_2011_AFM}
S.~Schmidt, P.~A.~L. Fernandes, B.~G. De~Geest, A.~G. Delcea, M.~Skirtach,
  H.~M{\"o}hwald and A.~Fery, \emph{Adv. Funct. Mater.}, 2011, \textbf{21},
  1411--1418\relax
\mciteBstWouldAddEndPuncttrue
\mciteSetBstMidEndSepPunct{\mcitedefaultmidpunct}
{\mcitedefaultendpunct}{\mcitedefaultseppunct}\relax
\EndOfBibitem
\bibitem[Sukhorukov \emph{et~al.}(1999)Sukhorukov, Brumen, Donath, and
  M{\"o}hwald]{Sukhorukov_1999_JPCB}
G.~B. Sukhorukov, M.~Brumen, E.~Donath and H.~M{\"o}hwald, \emph{J. Phys. Chem.
  B}, 1999, \textbf{103}, 6434--6440\relax
\mciteBstWouldAddEndPuncttrue
\mciteSetBstMidEndSepPunct{\mcitedefaultmidpunct}
{\mcitedefaultendpunct}{\mcitedefaultseppunct}\relax
\EndOfBibitem
\bibitem[Gao \emph{et~al.}(2001)Gao, Donath, Moya, Dudnik, and
  M{\"o}hwald]{Gao_2001_EPJE}
C.~Gao, E.~Donath, S.~Moya, V.~Dudnik and H.~M{\"o}hwald, \emph{Eur. Phys. J.
  E}, 2001, \textbf{5}, 21--27\relax
\mciteBstWouldAddEndPuncttrue
\mciteSetBstMidEndSepPunct{\mcitedefaultmidpunct}
{\mcitedefaultendpunct}{\mcitedefaultseppunct}\relax
\EndOfBibitem
\bibitem[Halo{\v z}an \emph{et~al.}(2005)Halo{\v z}an, D{\' e}jugnat, Brumen,
  and Sukhorukov]{Halozan_2005_JCIM}
D.~Halo{\v z}an, C.~D{\' e}jugnat, M.~Brumen and G.~B. Sukhorukov, \emph{J.
  Chem. Inf. Model.}, 2005, \textbf{45}, 1589--1592\relax
\mciteBstWouldAddEndPuncttrue
\mciteSetBstMidEndSepPunct{\mcitedefaultmidpunct}
{\mcitedefaultendpunct}{\mcitedefaultseppunct}\relax
\EndOfBibitem
\bibitem[Kim \emph{et~al.}(2005)Kim, Fan, Lebedeva, and
  Vinogradova]{Vinogradova_2005_Macromol}
B.-S. Kim, T.-H. Fan, O.~V. Lebedeva and O.~I. Vinogradova, \emph{Macromol.},
  2005, \textbf{38}, 8066--8070\relax
\mciteBstWouldAddEndPuncttrue
\mciteSetBstMidEndSepPunct{\mcitedefaultmidpunct}
{\mcitedefaultendpunct}{\mcitedefaultseppunct}\relax
\EndOfBibitem
\bibitem[Halo{\v z}an \emph{et~al.}(2007)Halo{\v z}an, Sukhorukov, Brumen,
  Donath, and M{\" o}hwald]{Halozan_2007_ACS}
D.~Halo{\v z}an, G.~B. Sukhorukov, M.~Brumen, E.~Donath and H.~M{\" o}hwald,
  \emph{Acta Chim. Slov.}, 2007, \textbf{54}, 598--604\relax
\mciteBstWouldAddEndPuncttrue
\mciteSetBstMidEndSepPunct{\mcitedefaultmidpunct}
{\mcitedefaultendpunct}{\mcitedefaultseppunct}\relax
\EndOfBibitem
\bibitem[Stukan \emph{et~al.}(2006)Stukan, Lobaskin, Holm, and
  Vinogradova]{Stukan_2006_PRE}
M.~R. Stukan, V.~Lobaskin, C.~Holm and O.~I. Vinogradova, \emph{Phys. Rev. E},
  2006, \textbf{73}, 021801\relax
\mciteBstWouldAddEndPuncttrue
\mciteSetBstMidEndSepPunct{\mcitedefaultmidpunct}
{\mcitedefaultendpunct}{\mcitedefaultseppunct}\relax
\EndOfBibitem
\bibitem[Tsekov and Vinogradova(2007)]{Vinogradova_2007_JCP}
R.~Tsekov and O.~I. Vinogradova, \emph{J. Chem. Phys.}, 2007, \textbf{126},
  094901\relax
\mciteBstWouldAddEndPuncttrue
\mciteSetBstMidEndSepPunct{\mcitedefaultmidpunct}
{\mcitedefaultendpunct}{\mcitedefaultseppunct}\relax
\EndOfBibitem
\bibitem[Tsekov \emph{et~al.}(2008)Tsekov, Stukan, and
  Vinogradova]{Vinogradova_2008_JCP}
R.~Tsekov, M.~R. Stukan and O.~I. Vinogradova, \emph{J. Chem. Phys.}, 2008,
  \textbf{129}, 244707\relax
\mciteBstWouldAddEndPuncttrue
\mciteSetBstMidEndSepPunct{\mcitedefaultmidpunct}
{\mcitedefaultendpunct}{\mcitedefaultseppunct}\relax
\EndOfBibitem
\bibitem[Lobaskin \emph{et~al.}(2012)Lobaskin, Bogdanov, and
  Vinogradova]{Lobaskin_2012_SM}
V.~Lobaskin, A.~N. Bogdanov and O.~I. Vinogradova, \emph{Soft Matter}, 2012,
  \textbf{8}, 9428--9435\relax
\mciteBstWouldAddEndPuncttrue
\mciteSetBstMidEndSepPunct{\mcitedefaultmidpunct}
{\mcitedefaultendpunct}{\mcitedefaultseppunct}\relax
\EndOfBibitem
\bibitem[Cordova \emph{et~al.}(2003)Cordova, Deserno, Gelbart, and
  Ben-Shaul]{Cordova_2003_BiophysJ}
A.~Cordova, M.~Deserno, W.~M. Gelbart and A.~Ben-Shaul, \emph{Biophys J.},
  2003, \textbf{85}, 70--74\relax
\mciteBstWouldAddEndPuncttrue
\mciteSetBstMidEndSepPunct{\mcitedefaultmidpunct}
{\mcitedefaultendpunct}{\mcitedefaultseppunct}\relax
\EndOfBibitem
\bibitem[Tang and Denton(2014)]{Tang_2014_PCCP}
Q.~Tang and A.~R. Denton, \emph{Phys. Chem. Chem. Phys.}, 2014, \textbf{16},
  20924--20931\relax
\mciteBstWouldAddEndPuncttrue
\mciteSetBstMidEndSepPunct{\mcitedefaultmidpunct}
{\mcitedefaultendpunct}{\mcitedefaultseppunct}\relax
\EndOfBibitem
\bibitem[Oosawa(1971)]{Oosawa_1971}
F.~Oosawa, \emph{Polyelectrolytes}, Dekker, New York, 1971\relax
\mciteBstWouldAddEndPuncttrue
\mciteSetBstMidEndSepPunct{\mcitedefaultmidpunct}
{\mcitedefaultendpunct}{\mcitedefaultseppunct}\relax
\EndOfBibitem
\bibitem[Parthasarathy and Klingenberg(1996)]{Parthasarathy_1996_MSE}
M.~Parthasarathy and D.~J. Klingenberg, \emph{Mater. Sci. Eng.}, 1996,
  \textbf{17}, 57--103\relax
\mciteBstWouldAddEndPuncttrue
\mciteSetBstMidEndSepPunct{\mcitedefaultmidpunct}
{\mcitedefaultendpunct}{\mcitedefaultseppunct}\relax
\EndOfBibitem
\bibitem[Mohanty \emph{et~al.}(2012)Mohanty, Yethiraj, and
  Schurtenberger]{Mohanty_2012_SM}
P.~S. Mohanty, A.~Yethiraj and P.~Schurtenberger, \emph{Soft Matter}, 2012,
  \textbf{8}, 10819--10822\relax
\mciteBstWouldAddEndPuncttrue
\mciteSetBstMidEndSepPunct{\mcitedefaultmidpunct}
{\mcitedefaultendpunct}{\mcitedefaultseppunct}\relax
\EndOfBibitem
\bibitem[Deserno and Holm(2001)]{Deserno-Holm_2001}
M.~Deserno and C.~Holm, in \emph{Electrostatic Effects in Soft Matter and
  Biophysics}, ed. C.~Holm, P.~K{\'e}kicheff and R.~Podgornik, Kluwer,
  Dordrecht, 2001, vol.~46, pp. 27--52\relax
\mciteBstWouldAddEndPuncttrue
\mciteSetBstMidEndSepPunct{\mcitedefaultmidpunct}
{\mcitedefaultendpunct}{\mcitedefaultseppunct}\relax
\EndOfBibitem
\bibitem[L{\"o}wen \emph{et~al.}(1992)L{\"o}wen, Madden, and
  Hansen]{Lowen_1992_PRL}
H.~L{\"o}wen, P.~A. Madden and J.-P. Hansen, \emph{Phys. Rev. Lett.}, 1992,
  \textbf{68}, 1081--1085\relax
\mciteBstWouldAddEndPuncttrue
\mciteSetBstMidEndSepPunct{\mcitedefaultmidpunct}
{\mcitedefaultendpunct}{\mcitedefaultseppunct}\relax
\EndOfBibitem
\bibitem[L{\"o}wen \emph{et~al.}(1993)L{\"o}wen, Madden, and
  Hansen]{Lowen_1993_JCP}
H.~L{\"o}wen, P.~A. Madden and J.-P. Hansen, \emph{J. Chem. Phys.}, 1993,
  \textbf{98}, 3275--3289\relax
\mciteBstWouldAddEndPuncttrue
\mciteSetBstMidEndSepPunct{\mcitedefaultmidpunct}
{\mcitedefaultendpunct}{\mcitedefaultseppunct}\relax
\EndOfBibitem
\bibitem[Dobnikar \emph{et~al.}(2003)Dobnikar, Chen, Rzehak, and von
  Gr{\"u}nberg]{Dobnikar_2003_JCP}
J.~Dobnikar, Y.~Chen, R.~Rzehak and H.~H. von Gr{\"u}nberg, \emph{J. Chem.
  Phys.}, 2003, \textbf{119}, 4971--4985\relax
\mciteBstWouldAddEndPuncttrue
\mciteSetBstMidEndSepPunct{\mcitedefaultmidpunct}
{\mcitedefaultendpunct}{\mcitedefaultseppunct}\relax
\EndOfBibitem
\bibitem[Dobnikar \emph{et~al.}(2003)Dobnikar, Chen, Rzehak, and von
  Gr{\"u}nberg]{Dobnikar_2003_JPCM}
J.~Dobnikar, Y.~Chen, R.~Rzehak and H.~H. von Gr{\"u}nberg, \emph{J. Phys.:
  Condens. Matter}, 2003, \textbf{15}, S263--S268\relax
\mciteBstWouldAddEndPuncttrue
\mciteSetBstMidEndSepPunct{\mcitedefaultmidpunct}
{\mcitedefaultendpunct}{\mcitedefaultseppunct}\relax
\EndOfBibitem
\bibitem[Hallez \emph{et~al.}(2014)Hallez, Diatta, and
  Meireles]{Hallez_2014_Langmuir}
Y.~Hallez, J.~Diatta and M.~Meireles, \emph{Langmuir}, 2014, \textbf{30},
  6721--6729\relax
\mciteBstWouldAddEndPuncttrue
\mciteSetBstMidEndSepPunct{\mcitedefaultmidpunct}
{\mcitedefaultendpunct}{\mcitedefaultseppunct}\relax
\EndOfBibitem
\bibitem[Marcus(1955)]{Marcus_1955_JCP}
R.~A. Marcus, \emph{J. Chem. Phys.}, 1955, \textbf{23}, 1057--1068\relax
\mciteBstWouldAddEndPuncttrue
\mciteSetBstMidEndSepPunct{\mcitedefaultmidpunct}
{\mcitedefaultendpunct}{\mcitedefaultseppunct}\relax
\EndOfBibitem
\bibitem[Wennerstr{\"o}m \emph{et~al.}(1982)Wennerstr{\"o}m, J{\"o}nsson, and
  Linse]{Wennerstrom_1982_JCP}
H.~Wennerstr{\"o}m, B.~J{\"o}nsson and P.~Linse, \emph{J. Chem. Phys.}, 1982,
  \textbf{76}, 4665--4670\relax
\mciteBstWouldAddEndPuncttrue
\mciteSetBstMidEndSepPunct{\mcitedefaultmidpunct}
{\mcitedefaultendpunct}{\mcitedefaultseppunct}\relax
\EndOfBibitem
\bibitem[Denton(2010)]{Denton_2010_JPCM}
A.~R. Denton, \emph{J. Phys.: Condens. Matter}, 2010, \textbf{22},
  364108--1--8\relax
\mciteBstWouldAddEndPuncttrue
\mciteSetBstMidEndSepPunct{\mcitedefaultmidpunct}
{\mcitedefaultendpunct}{\mcitedefaultseppunct}\relax
\EndOfBibitem
\bibitem[Flory(1953)]{Flory1953}
P.~J. Flory, \emph{Principles of Polymer Chemistry}, Cornell University Press,
  Ithaca, 1953\relax
\mciteBstWouldAddEndPuncttrue
\mciteSetBstMidEndSepPunct{\mcitedefaultmidpunct}
{\mcitedefaultendpunct}{\mcitedefaultseppunct}\relax
\EndOfBibitem
\bibitem[Forsman(2004)]{Forsman_2004_JPCB}
J.~Forsman, \emph{J. Phys. Chem. B}, 2004, \textbf{108}, 9236--9245\relax
\mciteBstWouldAddEndPuncttrue
\mciteSetBstMidEndSepPunct{\mcitedefaultmidpunct}
{\mcitedefaultendpunct}{\mcitedefaultseppunct}\relax
\EndOfBibitem
\end{mcitethebibliography}

\providecommand*{\mcitethebibliography}{\thebibliography}
\csname @ifundefined\endcsname{endmcitethebibliography}
{\let\endmcitethebibliography\endthebibliography}{}

\end{document}